\documentclass[aps,prl,reprint, superscriptaddress,showpacs,longbibliography]{revtex4-1}
\usepackage[english]{babel}
\usepackage{amsmath}
\usepackage{amsfonts}
\usepackage{amssymb}
\usepackage{graphicx}
\usepackage{bm}
\usepackage{bbm}

\begin{document}

\title{Fast charge sensing of a cavity-coupled double quantum dot using a Josephson parametric amplifier}
\author{J. Stehlik}
\affiliation{Department of Physics, Princeton University, Princeton, NJ 08544, USA}
\author{Y.-Y. Liu}
\affiliation{Department of Physics, Princeton University, Princeton, NJ 08544, USA}
\author{C. M. Quintana}
\affiliation{Department of Physics, Princeton University, Princeton, NJ 08544, USA}
\affiliation{Department of Physics, University of California, Santa Barbara, CA 93106, USA}
\author{C. Eichler}
\affiliation{Department of Physics, Princeton University, Princeton, NJ 08544, USA}
\author{T. R. Hartke}
\affiliation{Department of Physics, Princeton University, Princeton, NJ 08544, USA}
\author{J. R. Petta}
\affiliation{Department of Physics, Princeton University, Princeton, NJ 08544, USA}
\affiliation{Department of Physics, University of California, Santa Barbara, CA 93106, USA}
% 03.67.Lx - Quantum computation architectures and implementations
% 73.63.Kv - electronic transport in QD
% 85.35.Gv - single electron devices

\begin{abstract}
We demonstrate fast readout of a double quantum dot (DQD) that is coupled to a superconducting resonator.  Utilizing parametric amplification in a nonlinear operational mode, we improve the signal-to-noise ratio (SNR) by a factor of 2000 compared to the situation with the parametric amplifier turned off.   With an integration time of 400 ns we achieve a SNR of 76.  By studying SNR as a function of the integration time we extract an equivalent charge sensitivity of $8 \times 10^{-5} \:\:e/\sqrt{\rm Hz}$.  The high SNR allows us to acquire a DQD charge stability diagram in just 20 ms. At such a high data rate, it is possible to acquire charge stability diagrams in a live ``video-mode," enabling real time tuning of the DQD confinement potential.
\end{abstract}

%Utilizing parametric amplification we improve the signal-to-noise ratio (SNR) by a factor of 2000 relative to a setup that only uses a standard high electron mobility transistor (HEMT) amplifier.

\maketitle

\section{I. Introduction}

Conventional charge transport experiments directly measure the electrical conductance or current flow through a sample. Instead of measuring the rate at which electrons flow through a sample, it is sometimes desirable to have direct access to the charge degree of freedom using sensitive electrometers. For example, charge sensing can be used for real-time counting of electrons \cite{currentMeasurement}, imaging of current flow through nanostructures such as quantum point contacts \cite{Topinka29092000}, imaging electron-hole puddles in graphene \cite{graphenePuddles}, and single electron capacitance spectroscopy of quantum Hall states \cite{Ashoori1992,Dial}.
Charge detection is also a necessary ingredient for quantum computation using electrons trapped in gate defined quantum dots \cite{HansonRev,RevModPhys.75.1}. For spin qubits, a process called spin-to-charge conversion is used for readout \cite{PauliBlockade,ACJohnsonSpinBlocakde}. Charge sensing is also needed to accurately measure the charge stability diagram of double quantum dots (DQD) and determine the absolute electronic occupancy. Sensing in quantum dots has traditionally been performed using quantum point contacts (QPC) or single electron transistors (SET) \cite{QPC1,DevoretSchoelkopfreview,RimbergNatureSET}.

One method to improve the charge sensing bandwidth involves embedding the QPC or SET in a resonant tank circuit and performing rf-reflectometry \cite{Schoelkopf_RFSET,Kontos_PhysRevB,karlRF}. Reflectometry experiments have so far utilized standard cryogenic low-noise amplifiers \cite{Weinreb}. The increased data acquisition rate of rf-based measurements reduces the susceptibility to $1/f$ noise and enables single-shot spin readout \cite{MarcusSingleShot}.

While rf-reflectometry has allowed charge detection times as short as 100 ns with a signal-to-noise ratio (SNR) $\sim 10$  \cite{MarcusFastSense,ReillyFastSensing}, recent progress in superconducting quantum circuits opens a path forward for further improvements.  The application of quantum limited Josephson parametric amplifiers (JPAs) \cite{paraampMisc,paraampMisc2,LehnertJPA,jpaRevSciInst,paraampMisc4,paraampMisc5,paraampMisc6} has accelerated progress in the field of superconducting qubits.  These amplifiers are nearly quantum limited, meaning they add the minimum amount of noise dictated by the uncertainty relation \cite{CavesPRD1982,CavesPRA2012,LehnertJPA}, and have enabled high fidelity single shot readout of qubit states \cite{VijaySingleShot} as well as squeezed light experiments \cite{Squeezed1,LehnertJPA,EichlerPRL2011}.
%These amplifiers confer two large benefits to the readout chain.  They are quantum limited, meaning they add the minimum amount of noise dictated by the uncertainty relation \cite{CavesPRD1982,CavesPRA2012,LehnertJPA}, and since they are nearly dissipationless they can be placed on the mixing chamber of a dilution refrigerator, and thus extremely close to the sample. The close proximity of the sample to the amplifier minimizes the all important loss before the first stage of amplification \cite{PozarNoiseChapter}.

%combine some of the best elements of superconducting quantum circuits with semiconductor quantum dots. We
In this paper we  demonstrate JPA-assisted readout of a DQD that is coupled to a high frequency superconducting cavity.  We achieve a factor of 2000 improvement in the SNR by operating the JPA in a nonlinear regime, in which the average readout signal experiences a larger amplification than the incoming noise. With an integration time of just 400 ns, we achieve a SNR of 76.  The large SNR allows us to efficiently map out the DQD charge stability diagram in just 20 ms using a dual gate-voltage rastering scheme. Such dramatic improvements enable real-time ``video-mode" tuning of a DQD device.  The faster device optimization that JPA assisted readout allows will become crucial as quantum dot architectures with more control parameters are developed \cite{SachrajdaTQD,TaruchaAPL,PettaAwschalomScience}.

\section{II. Josephson Parametric Amplifier Assisted Readout of a Cavity-Coupled Semiconductor Double Quantum Dot}

Our device architecture is schematically illustrated in Fig.\ 1(a).  The experiment consists of three main components: a DQD formed in an InAs nanowire, a Nb superconducting cavity to which the DQD is coupled, and a JPA that is used to amplify the output field of the cavity. In the following sections we describe each of these components in detail.

\subsection{A. InAs Double Quantum Dot}

The DQD is fabricated using a bottom-gated architecture as shown in Fig.\ 1(b) \cite{StefanNP}.  We place a single InAs nanowire of nominally 50 nm diameter \cite{SchroerNano} across five depletion gates \cite{FasthInAs}.  We label the gates $V_{\rm LW}$, $V_{\rm L}$, $V_{\rm M}$, $V_{\rm R}$, and $V_{\rm RW}$.  A double-well potential is formed by selectively depleting the nanowire using the bottom gates \cite{FasthInAs}.  Here $V_{\rm LW}$ and $V_{\rm RW}$ control the left and right barriers, $V_{\rm L}$ and $V_{\rm R}$ control $\mu_{\rm L}$ and $\mu_{\rm R}$, the electrochemical potential of each dot, and $V_{\rm M}$ controls the interdot tunneling rate \cite{RevModPhys.75.1}.  After passivating the nanowire using an ammonium polysulfide etch \cite{SulfurEtch} we deposit Ti/Au contacts to form the source and drain electrodes. The source is connected to a voltage anti-node of a half-wavelength superconducting resonator and the drain is connected to the resonator ground plane \cite{KarlFancy}.

\subsection{B. Superconducting Cavity}

%========================
% Details:
% Per Yinu's analysis, cap is 3.6 aF
% 250 nm of thermal oxide
The cavity is implemented as a coplanar half-wavelength superconducting resonator with a resonance frequency $f_{\rm c} = 7881$ MHz, see Fig.\ 1(b).  A 50 nm thick Nb film is sputter deposited on a high resistivity silicon substrate with a 250 nm thick layer of thermally grown $\mathrm{SiO}_2$.  The input and output capacitors result in input and output coupling rates $\kappa_{\rm in} / 2\pi \approx \kappa_{\rm out} / 2\pi \approx 0.39$ MHz.  Internal radiation losses result in a total cavity decay rate $\kappa_{\rm tot}/ 2\pi = 2.6$ MHz equivalent to a loaded quality factor $Q_{\rm c} \sim 3000$.  The source contact of the nanowire is connected to a voltage anti-node of the resonator, resulting in a large electric dipole coupling between the DQD and the resonator \cite{cavityDQDWallraff1,KarlFancy,KontosCoupling}.  We probe the hybrid system by applying a tone $f_{\rm in} = f_{\rm c}$ of power $P_{\rm in} \approx -110$ dBm to the input port of the cavity. Charge dynamics in the DQD change the effective admittance of the microwave cavity leading to a measurable dispersive shift of the resonance frequency \cite{cavityDQDWallraff1,KarlFancy,KontosCoupling}.
%Charge dynamics in the DQD result in an admittance that loads the microwave cavity and leads to a dispersive shift of the cavity resonance frequency \cite{cavityDQDWallraff1,KarlFancy,KontosCoupling}.

\subsection{C. Josephson Parametric Amplifier}

%=================
% Coupling capacitance = 39.12 fF
% Simulated in maxwell file: cap100
% Kappa arrived at by formula on page 97 of Eichler's thesis is
% 100 MHz.

\begin{figure}[t]
\begin{center}
\includegraphics[width=\columnwidth]{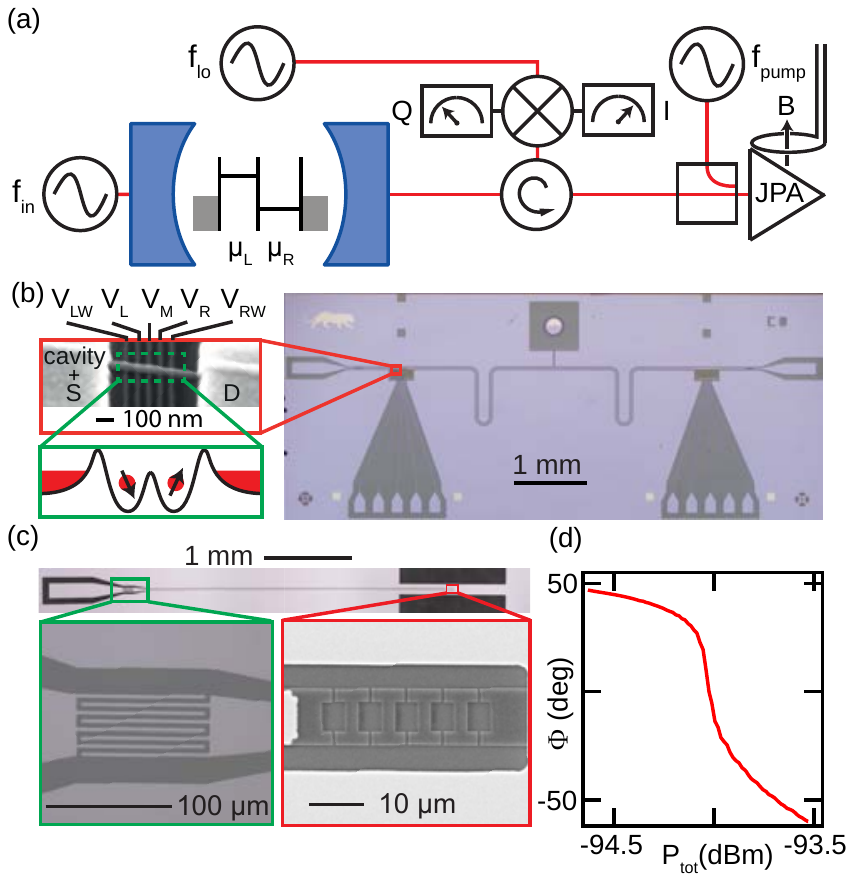}
\caption{(Color online) (a) Diagram of the experimental setup.  A tone $f_{\rm in}$ is used to probe a cavity-coupled DQD.  The output field of the cavity is amplified using a JPA and then demodulated into the $I$ and $Q$ quadratures.  The amplification band of the JPA is flux-tuned using a small coil that creates a magnetic field $B$. (b) Illustration of the DQD-cavity system.  A single nanowire is placed on top of 5 depletion gates. Negative voltages on the gates establish a double-well potential, forming the DQD. (c) Optical micrograph of the JPA, which is implemented as a quarter-wavelength resonator shunted by an array of 5 SQUIDs.  The resonator is coupled to the input port using an interdigitated planar capacitor (outlined in green).  A scanning electron micrograph showing the SQUID array is outlined in red.  (d) Phase $\Phi$ of the signal reflected off of the JPA as a function of the total incident power $P_{\rm tot}$.  We operate in phase-sensitive mode with $f_{\rm in} = f_{\rm pump}$ and use the strong phase dependence to translate small differences in the cavity output field to large phase shifts of the microwaves reflected off of the JPA.}
\label{fig1}
\end{center}	
\vspace{-0.6cm}
\end{figure}

As shown in Fig.\ 1(c) the JPA is implemented as a quarter-wavelength resonator etched from a 200 nm Nb layer on top of a sapphire substrate \cite{LehnertJPA,EichlerPRL2011}.  The input port has a coupling rate $\kappa_{\rm JPA}$ and is formed using an interdigitated capacitor geometry.  The resonator is shunted by a series array of $M = 5$ superconducting quantum interference devices (SQUIDs).  The SQUIDs act as a non-linear inductor and allow for amplification of the signal field via the pump field resulting in a Lorentzian gain profile with nearly constant gain-bandwidth product: $\sqrt{G} B_{\rm w} \propto \kappa_{\rm JPA} / 2\pi$ \cite{LehnertJPA}.  Here $G$ is the JPA power gain (which can be tuned using the applied pump power $P_{\rm pump}$), while $B_{\rm w}$ is the 3 dB bandwidth.  The dynamic range of the amplifier is set by the strength of the pump field required for a given gain and can be improved by increasing the number of SQUIDs \cite{dynamicRange}.  We choose $M=5$ with each SQUID having maximal Josephson energy $E_{\rm J }/h = 7 \pm 0.5$ THz and design our coupling capacitance to be $40$ fF for a target $\kappa_{\rm JPA} / 2 \pi \sim 100$ MHz. A detailed treatment of the design considerations of this class of amplifiers is given in Ref.\ \cite{dynamicRange}.
%$\kappa_{\rm JPA}/2\pi$ cannot be increased without limit, as the ac current through the SQUIDs will exceed the critical current $I_{\rm c}$.  The limitations imposed can be partially reduced by using an array of SQUIDs \cite{dynamicRange}.

We use the amplifier in phase-sensitive mode  \cite{LehnertJPA,Riste2012} and set $f_{\rm pump} = f_{\rm in} = 7881$ MHz, which is the resonance frequency of our cavity.  The JPA is tuned such that the reflected phase $\Phi$ is highly sensitive to the total incident power $P_{\rm tot}$.  Here $P_{\rm tot}$ is a combination of the pump power $P_{\rm pump}$ and the power of the cavity output field, $P_{\rm out}$. Figure 1(d) shows $\Phi$ as a function of $P_{\rm tot}$.  We set $P_{\rm pump} = -94$ dBm, which maximizes $d\Phi / dP_{\rm tot}$. Adding the cavity output field to the pump modifies the total power incident on the JPA.  This will result in large differences  of the phase of the reflected pump signal.  Thus in this measurement mode the small cavity output field modulates the phase of the much stronger reflected pump tone, leading to amplification.

We further improve the SNR of our readout chain by driving the JPA beyond the regime where it amplifies linearly.  We choose the power and the phase of the cavity drive field such that the dispersive shift changes the total power at the JPA from $P_{\rm tot} \approx -93.9$ dBm to $P_{\rm tot} \approx -94.1$ dBm \cite{SOM}.  As such, the dispersive shift will change $P_{\rm tot}$ from above the region of maximum $d\Phi / dP_{\rm tot}$ to below.  This means that while the dispersive shift produces a large change in $\Phi$, any noise fluctuations will invariably occur in regions with smaller $d\Phi/dP_{\rm tot}$ and be amplified less \cite{SOM}.  The bias point of the JPA is thus chosen such that its amplification characteristics fall between those of a linear amplifier and a bifurcation amplifier \cite{jpaRevSciInst}.
%is thus chosen such that its amplification characteristics fall between those of a linear amplifier and a bifurcation amplifier.
%For qubit state discrimination a further extension of this  technique leads to bifurcated readout \cite{jpaRevSciInst}.
% This effectively suppresses the signal coming from noise fluctuations,  Outside of this region $d\Phi/dP_{\rm tot}$ is significantly smaller, and as such fluctuations will not affect the readout signal as much, while the dispersive shift results in a very large phas shift \cite{SOM}.  Our readout strategy is thus a hybrid in between linear amplification chain and a bifurcated readout \cite{jpaRevSciInst}.
% the much stronger pump tone acts as a carrier, which is phase modulated by the small cavity signal using the JPA.

The signal exiting the JPA is further amplified using a high electron mobility transistor (HEMT) amplifier at the 3 K stage of the dilution refrigerator and room temperature amplifiers \cite{SOM}. We can turn off the JPA by simply removing the pump tone.  Here the signal from the cavity reflects off of the JPA with unity gain.  In both cases we demodulate the signal using a heterodyne detection setup.  We first downconvert the signal to 12.5 MHz by mixing it with a local oscillator tone $f_{\rm lo} = f_{\rm in}+12.5$ MHz. It is then digitized using a FPGA-based digital homodyne stage to recover the $I$ and $Q$ quadrature amplitudes \cite{SOM}.

\section{III. Results}

We now highlight the dramatic enhancement in the data acquisition rate that is enabled by the JPA. We first demonstrate rapid measurements of a large-scale DQD charge stability diagram using a JPA. The SNR values obtained with and without the JPA are then compared, indicating a factor of 2000 improvement with the JPA. Lastly, we demonstrate video-mode acquisition of charge stability diagrams using a rastered sweep of two gate voltages, resulting in data display that is similar to a cathode ray tube and of a comparable refresh rate.

\subsection{A. Double Quantum Dot Charge Stability Diagram}

A large scale charge stability diagram is shown in Fig.\ 2, where we plot the normalized $Q$ quadrature amplitude as a function of $V_{\rm L}$ and $V_{\rm R}$. These data are acquired by measuring the total reflected field from the JPA with the JPA pump tone on.  With the DQD in Coulomb blockade, the resonator is insensitive to the DQD, resulting in large regions in the stability diagram with uniform amplitude and phase. Following previous work, we normalize the amplitude $A = I+i Q$, such that $\left|A \right| = 1$ and $ \arg \left(A\right) = 1$ in these charge stability islands \cite{KarlFancy}.  With this normalization the sensitivity of the $Q$ quadrature to charge transitions is maximized and its mean value in Coulomb blockade is $<$$Q$$>$ = 0.
%Our normalization convention means that the $Q$ quadrature will be most sensitive to charge transitions and have a mean value of 0 when the DQD is in Coulomb blockade.

\begin{figure}
\begin{center}
\includegraphics[width=\columnwidth]{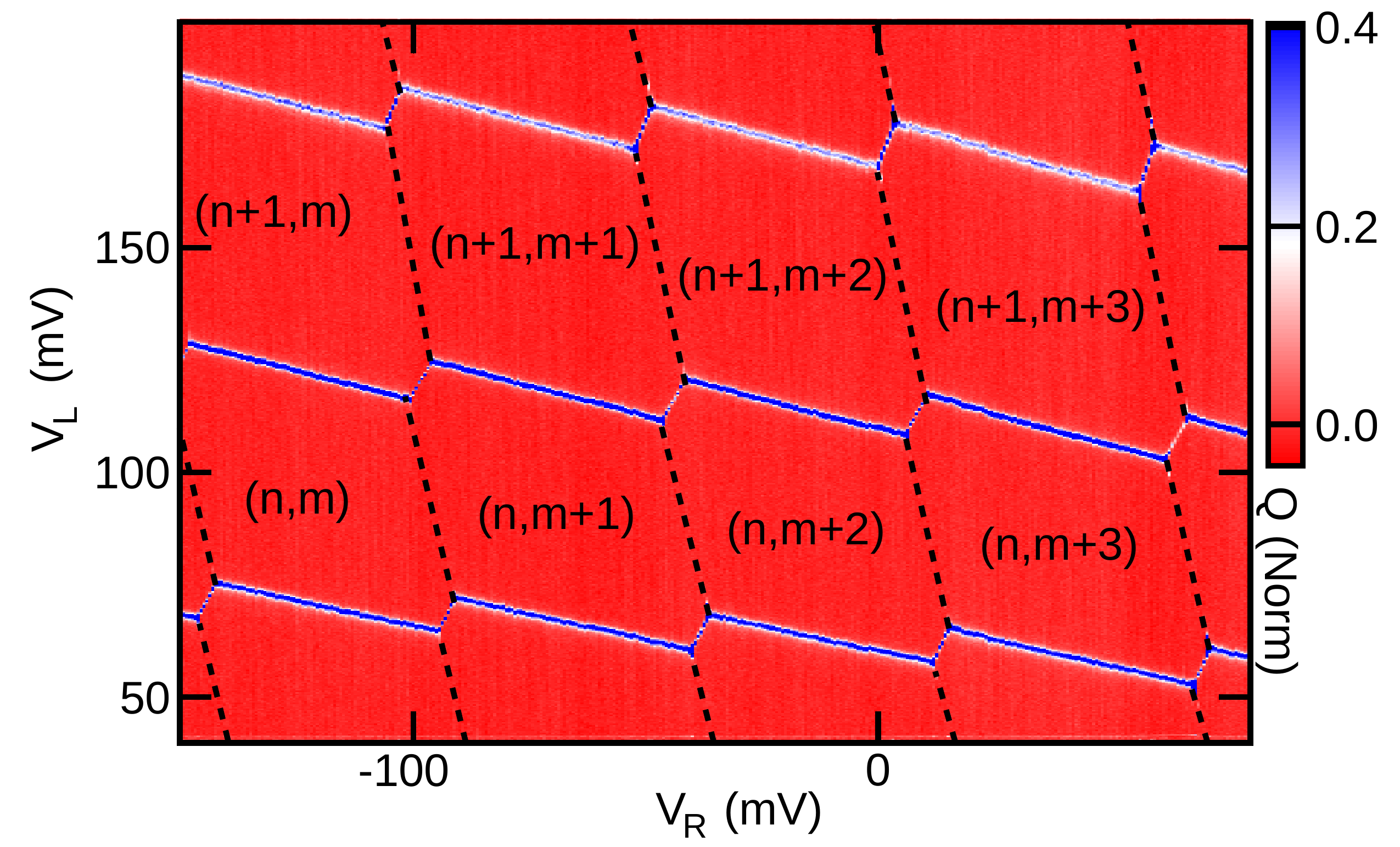}
\caption{(Color online)  DQD charge stability diagram extracted by plotting the normalized $Q$ quadrature amplitude as a function of gate voltages $V_{\rm L}$ and $V_{\rm R}$.  The plot was taken with the JPA turned on and only took 2 minutes to acquire. Dashed lines are guides to the eye. }
\label{harm2}
\end{center}
\vspace{-0.6cm}
\end{figure}

When an electron can tunnel between the two dots (interdot charge transition) or on and off one of the dots (single dot charge transition), the transmitted field through the resonator will be affected, which we detect as a non-zero $Q$ quadrature amplitude.  As a result the data in Fig.\ 2 trace out the charge stability diagram of the DQD \cite{RevModPhys.75.1}.  The nearly horizontal lines are associated with the addition or removal of an electron from the left dot. The lines with positive slope correspond to interdot charge transitions where the total electron number is fixed, but an electron moves from one dot to the other.  Due to the large distance between the cavity center pin and the right tunnel barrier, right dot charge transitions are not visible. We can infer their location based on the positions of the interdot charging lines. We use the charging lines to label the electron states as $(n,m)$, where $n$ ($m$) is the number of electrons on the left (right) dot. For this data set, the device is configured in the many-electron regime, where $n$ $\sim$ $m$ $\sim$ 20.  We note that the data shown in Fig.\ 2 were acquired in just 2 minutes.  In comparison, acquiring similar quality data using standard low frequency conductance measurements of a quantum point contact would require approximately an hour.
%This therefore indicates a dramatic increase in the SNR.

\subsection{B. Signal-to-Noise Measurements}

\begin{figure}
\begin{center}
\includegraphics[width=\columnwidth]{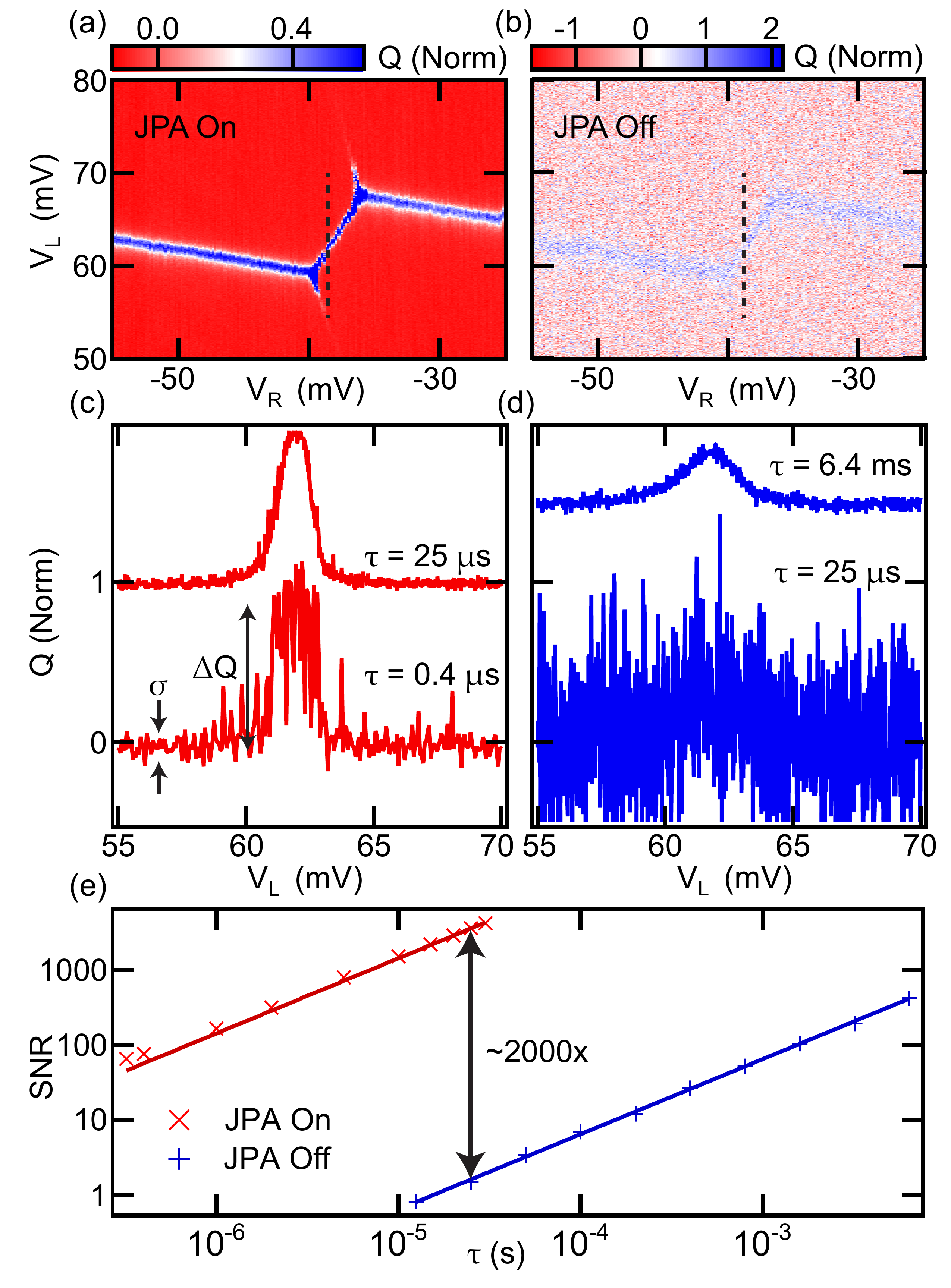}
\caption{(Color online) (a--b)  $Q$ quadrature amplitude plotted as a function of $V_{\rm L}$ and $V_{\rm R}$ near the $(n,m+1) \leftrightarrow (n-1,m+2)$ interdot charge transition.  (a) Data taken with the JPA on, and (b) with the JPA off.  For both cases each point of the colormap is integrated for $\tau = 42\ \mu \mathrm{s} $.  (c--d) Traces of the $Q$ quadrature amplitude showing the interdot charge transition for different per-point integration times $\tau$.  (c) Data taken with the JPA on (traces are offset by 1 for clarity), and (d) with the JPA off (traces are offset by 1.5 for clarity).   (e) Signal-to-noise ratio (SNR) for the interdot charge transition as a function of the integration time $\tau$.  Amplification with the JPA results in a factor of 2000 improvement in the SNR. Solid lines are fits to Eq.\ 2.}
\label{harm3}
\end{center}
\vspace{-0.6cm}
\end{figure}

To quantify the improvement in the SNR due to the JPA we investigate the sensing signal at an interdot charge transition. Interdot charge transitions are most relevant in quantum control experiments since interdot tunneling is spin selective and can be used to distinguish spin singlet and triplet states \cite{PettaSeminal,KarlFancy,SchroerParity,ACJohnsonSpinBlocakde}. Figures 3(a--b) show charge stability diagrams obtained near the $(n,m+1) \leftrightarrow (n-1,m+2)$ interdot charge transition. For each point of the colormap the $Q$ quadrature is averaged for $\tau = 42\:\:\mu \mathrm{s}$.  For a direct comparison, the data in Fig.\ 3(a) were acquired with the JPA turned on, while the data in Fig.\ 3(b) were acquired with the JPA turned off. The dramatic increase in the visibility of the charge transitions in Fig.\ 3(a) provides direct confirmation of increased SNR when the JPA is used.  For a quantitative comparison, we ramp the gate voltage $V_{\rm L}$ along the dashed lines shown in Figs.\ 3(a--b) with a 50 Hz repetition rate and record time traces of the resultant $Q$ quadrature amplitude.  To study the SNR as a function of the effective per point integration time $\tau$, we oversample at 12.5 MS/s and then apply an appropriately sized digital box-car filter \cite{SOM}.  The results for $\tau = 400$ ns and $\tau = 25 \ \mu \mathrm{s}$ with the JPA turned on are shown in Fig.\ 3(c).  Even with $\tau = 400 $ ns the interdot signal is clearly discernible. Following standard electronic references, we define the SNR as the ratio of the signal power to the noise power in the measured channel \cite{SNRDefinition}. Using this definition,
%\begin{equation}
%\mathrm{SNR} = 10 \mathrm{log}_{10} \left( \frac{\Delta Q^2}{\sigma^2 } %\right) \ \mathrm{dB},
%\end{equation}
\begin{equation}
\mathrm{SNR} =  \frac{\Delta Q^2}{\sigma^2 } ,
\end{equation}
where $ \Delta Q$ is the amplitude of the interdot charge transition signal and $\sigma$ is the root-mean-square amplitude of the noise of the $Q$ quadrature in Coulomb blockade [see Fig.\ 3(c)]. In the case of $\tau = 400$ ns, we obtain a SNR $\approx 76$.

As illustrated by the data in Fig.\ 3(d), the SNR is much worse when only the HEMT is used. With $\tau = 25 \ \mu \mathrm{s}$ the SNR = 1.4, meaning the charge transition is barely resolvable. The signal from the interdot charge transition can be recovered by increasing $\tau$. Increasing $\tau$ beyond $500 \:\: \mu$s is accomplished by averaging multiple time-traces \cite{SOM}. The result for $\tau = 6.4$ ms is shown in the upper trace of Fig.\ 3(d).

The SNR is plotted as a function of the effective integration time $\tau$ in Fig.\ 3(e) with the JPA on and off. In both cases the curves are very well fit with
%\begin{equation}
%\mathrm{SNR}(\tau) = 10 \log_{10} \left( \frac{\tau}{\tau_{\rm min}} \right) \: dB.
%\label{snrtau}
%\end{equation}
\begin{equation}
\mathrm{SNR}(\tau) =  \frac{\tau}{\tau_{\rm min}} .
\label{snrtau}
\end{equation}
Here $\tau_{\rm min}$ is a fitting parameter that corresponds to the minimum integration time required to achieve a SNR = 1.  With the JPA on, a best fit is obtained with $\tau^{\rm on}_{\rm min} = 7$ ns, while with the JPA off $\tau^{\rm off}_{\rm min} = 16 \ \mu$s, corresponding to a factor of 2000 improvement in the SNR. The linear dependence of the SNR on $\tau$ is expected and can be understood from the fact that $\tau$ sets the effective low-pass filter of the detection chain.  Decreasing $\tau$ increases the bandwidth and therefore the noise power, while the signal power will remain constant \cite{SNRDefinition}.

%===============================================
% Data extraction note: BW extracted from wave 16716, same JPA
% Configuration as was used for the data
% Igor multi-peak gives FWHM 3.77 MHz
%===============================================

We now estimate the minimum time required for single charge detection, in terms of both sensitivity and ultimate bandwidth. From the fitting described above the minimum integration time necessary for a SNR = 1 is $\tau_{\rm min}^{\rm on} = 7 $ ns. Since charge sensitivity is related to minimum integration time through: $S = e \sqrt{\tau_{\rm min}}$, our detector has an equivalent charge sensitivity of $8 \times 10^{-5} \ e / \sqrt{\rm Hz}$ \cite{ReillyFastSensing,MarcusSingleShot}.  The bandwidth of our detection chain is limited by the cavity linewidth and the JPA bandwidth.  The former is set by $\kappa_{\rm tot} / 2 \pi \approx 2.6$ MHz and can be extended by increasing the output coupling capacitance.  The latter is constrained by the JPA gain-bandwidth product. In our case we operated the JPA with a gain $G \approx 4000$ and a bandwidth of 3.8 MHz \cite{SOM}.  This limit can be circumvented by more advanced amplifier designs \cite{martinisHighBWJPA,EichlerNewAmps}.  The above means that we are bandwidth limited by the cavity to detection times longer than $\approx 400$ ns.

%However, achieving a 140 MHz bandwidth is not possible with the present setup. The two bandwidth limiting factors are the cavity linewidth and the JPA bandwidth.  The former is set by $\kappa_{\rm tot} / 2 \pi \approx 2.6$ MHz and can be extended by increasing the output coupling capacitance.  The latter is constrained by the JPA gain-bandwidth product. In our case we operated the JPA with a gain $G \approx 4000$ and a bandwidth of 3.8 MHz \cite{SOM}.  This limit can be circumvented by more advanced amplifier designs \cite{martinisHighBWJPA,EichlerNewAmps}.

\subsection{C. Video-Mode Data Acquisition}
\begin{figure}
\begin{center}
\includegraphics[width=\columnwidth]{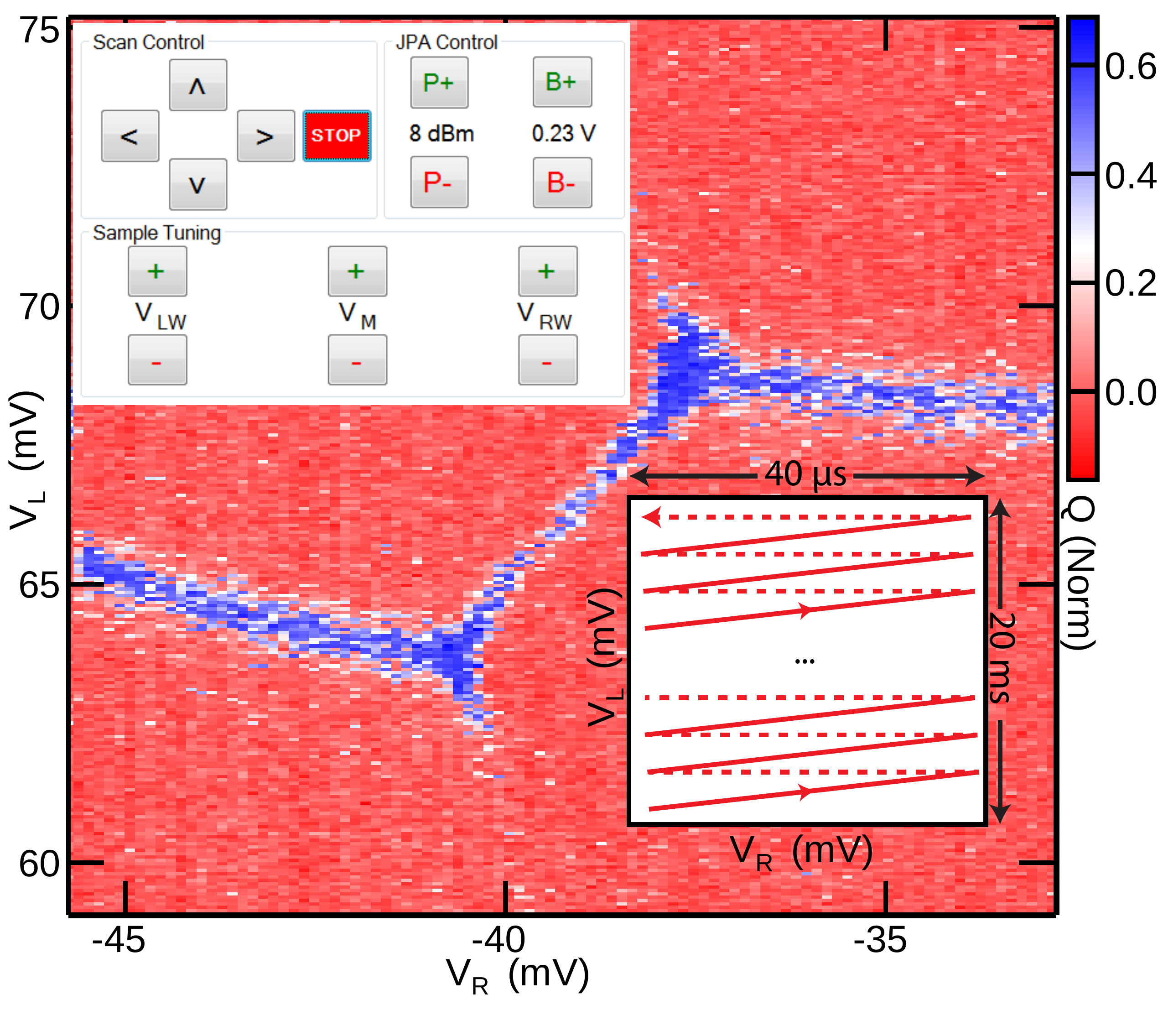}
\caption{(Color online) Normalized $Q$ quadrature amplitude as a function of $V_{\rm L}$ and $V_{\rm R}$.  The scan was acquired by rastering the gate voltages over a 20 ms period.  The rastering scheme is illustrated in the lower inset.  Starting in the lower left hand corner of the stability diagram, $V_{\rm R}$ is repeatedly ramped with a 40 $\mu$s period.  At the same time $V_{\rm L}$ is ramped up with a $20$ ms period. The rastering configuration is similar to cathode-ray-tube monitors.  The high SNR then allows for real-time tuning of the DQD.  This is accomplished using an intuitive user interface pictured in the upper inset. }
\label{harm4}
\end{center}
\vspace{-0.6cm}
\end{figure}

With the JPA on, the dramatic increase in the SNR allows for very fast acquisition of charge stability diagrams. For example with $\tau = 400$ ns a typical 100 $\times$ 100 point color map can be acquired in just 4 ms. To take advantage of this, we develop a dual gate-voltage rastering scheme.  This is illustrated in the lower inset of Fig.\ 4.  Starting in the lower left corner of the charge stability diagram, we repeatedly ramp $V_{\rm R}$ with a $40 \ \mu\mathrm{s}$ period.  At the same time we also ramp $V_{\rm L}$ with a $20$ ms period.  The resulting data set rasters over a two-dimensional voltage subspace within a 20 ms period.  During this time we record a single time trace of the measured quadrature  with $\tau = 400$ ns and reorder it to form a 2D image that can be displayed. A sample frame acquired using this approach is shown in Fig.\ 4.

As coupled quantum dot systems continue to scale in complexity, it will become increasingly important to rapidly obtain charge stability diagrams and optimize the relevant tunneling rates. Traditionally, gate voltages have been optimized through a slow iterative process that involves changing gate voltages and observing the resulting charge stability diagram.  The process is time consuming, in part due to the time necessary to acquire a new scan of the charge transition being tuned (often on the order of minutes).  We have built a simple user interface to control the experiment (see upper inset of Fig.\ 4) and use our rastering scheme to continuously update the charge transition in question.  A user can then simply click on a set of buttons to adjust electrostatic gate voltages, while the results of the parameter tuning are displayed in real-time. This gives immediate feedback and allows one to arrive at the optimal gate voltages with ease. A demonstration of this form of device tuning is captured in a video in the supplementary materials \cite{SOM}.

\section{IV. Conclusion and Outlook}

In conclusion, we have used a JPA for fast measurements of a cavity-coupled semiconductor DQD.  The addition of the JPA results in a factor of 2000 improvement in the SNR compared to traditional HEMT amplifier approaches. As a result, the minimum time to detect a single charge is limited by the linewidth of the cavity. We also develop a dual gate-voltage rastering scheme that allows us to measure a graph of a charge transition in 20 ms.  This enables real-time device tuning, which has the potential to greatly increase the rate at which complex quantum dot devices are optimized. Together with frequency multiplexing \cite{fMultiplexReilly,MartinisMultiplex}, JPA assisted readout may allow simultaneous fast readout of several DQDs.

We thank Thomas Hazard for helpful discussions and Drew Baden for assistance with FPGA programming. Research was supported by the Packard Foundation, ARO Grant No.\ W911NF-08-1-0189, DARPA QuEST Grant No.\ HR0011-09-1-0007, and the NSF (DMR-1409556 and DMR-1420541).

\bibliographystyle{apsrev4-1custom}

%

%\bibliography{SensingAPL}

\end{document}